\documentclass[12pt]{article}
\usepackage{amssymb,amsmath}

\newtheorem{them}{Theorem}[section]
\newtheorem{no}{Remark}[section]

\title{EXACT SELF-SIMILAR AND TWO-PHASE SOLUTIONS OF SYSTEMS OF 
SEMILINEAR PARABOLIC EQUATIONS}
\author{K. A. Volosov, V. G. Danilov, and A. M. Loginov
\\ Moscow. 
Institute of Electronics and Mathematics\\
contsam@dol.ru   }

\date{}

\begin{document}
\maketitle 
\setcounter{equation}{0}

\begin{sloppypar}

 \begin{abstract}
Interest to the tasks by bound with systems of the quasilinear
equations recently has revived.   
Therefore we have decided  to expose in electronic sort our
operation published in  
 Translated from Teoreticheskaya i Matematicheskaya Fizika, Vol.
101, No. 2,  
pp. 189-199, November, 1994. Original article submitted October
26, 1993. 

Exact single-wave and two-wave solutions of systems of equations of the 
Newell-Whitehead type are presented.  The Painleve test and calculations 
in the spirit of Hirota are used to construct these solutions.

\end{abstract}


\section*{INTRODUCTION}

In [1-4], methods were developed for constructing the explicit expressions 
that describe nonlinear waves (kinks) that are solutions of the semilinear
parabolic equations
\begin{equation}
 U_t-U_{xx}=F(U) ,
\label{1}
\end{equation} 
where $F(U)$  are quadratic and cubic polynomials.   
The solutions of $Eq. (0.1)$ considered in the cited studies satisfy boundary
conditions that
 are determined by the roots of the nonlinearity, i.e., solutions 
propagating in strips were constructed. Besides bounded solutions,
 solutions having a discontinuity of the second kind (so-called monsters) 
were considered.    In addotion, expression were obtained for
solutions that describe the interaction of kinks and also kinks 
in discontinuous solutions. At the present time, only two types of solution
 of $Eq. (0.1)$ represented by explicit expressions are known - so-called 
self-similar (single-phase, single-wave) solutions and solutions describing 
the interaction of self-similar waves (two-phase solutions) [1,4].
 The self-similar solutions have the form (for the notation, see below)
\begin{eqnarray}
&& \chi(\tau)= U(\frac{\varphi \exp{(\lambda\tau)}}{\psi \exp{(\lambda\tau)}}),\nonumber\\
&& \chi(\tau)= U(\frac{\varphi(\lambda\tau)}{\psi(\lambda\tau)}),
\label{2}
\end{eqnarray}
For the nonself-similar solutions describing the interaction of 
single-phase waves, there is a representation of the form
\begin{eqnarray}
&& \chi(\tau_1,\tau_2)= U(\frac{\varphi(\exp{(\lambda_1\tau_1)},\exp{(\lambda_2\tau_2),\tau_1,\tau_2)}}
{\psi(\exp{(\lambda_1\tau_1)},\exp{(\lambda_2\tau_2),\tau_1,\tau_2)}}),\nonumber\\
\label{3}
\end{eqnarray}
Here,$ U(z) $  is a function that is entire in some region $ \Omega\in C^{1} $;
 $\varphi$ and $\psi$ are polynomials; $\tau=x+pt + c $is the self-similar
 variable; $\lambda $ and $p$ are phase (related) constants; 
and $ c $ is an arbitrary constant. Similarly,$ \tau_i= x+p_{i} + c_{i},
 i=l, 2,$ where  $\lambda _i $ and $ p_i $ - are phase constants, 
$ c_i $ is an arbitrary constant, and the phases ,$ \tau_i $ are independent,
 i.e., a linear combination of them with integer coefficients does not vanish 
identically. We note that self-similar solutions of $ Eq. (0.1)$ are 
determined by the solution of the ordinary differential equation
\begin{eqnarray}
&&\lambda p\chi^{'}-{\lambda}^{2}\chi^{''}-F(\chi)=0.
\label{4}
\end{eqnarray}
It is shown in [3,4] that for equations of the form (0.4) the store
 of solutions of the form (0.2) and, hence, the set of constants $\lambda$ 
and $ p $ 
is very restricted. Similar results have also been obtained by means of
 the Painlev$\acute{e}$  test [5-7], the idea of which is to represent the solution 
of an equation of the form (1) by a Laurent series in the neighborhood of 
a moving pole, i.e., to expand the required function in powers of $\tau$, 
the self-similar variable.
 The Painlev$\acute{e}$ test, applied to various equations of the form (0.1),
 showed that the set of possible velocities for self-similar solutions
 represented by a Laurent series is, as a rule, small and can be uniquely 
determined. In addition, for each concrete value of the velocity $ p $
obtained by means of the Painlev$\acute{e}$  test the Laurent series can
 uniquely, 
up to a shift constant, determine a function of the form (0.2) .
  Therefore, in the case of a positive Painlev$\acute{e}$  test,
 depending on its 
results, it is natural to assume the existence for equations of the form 
(0.1) of solutions in the class of functions defined by the expressions 
(0.2)-(0.3). In this case, one can substitute in the original equation the 
general
form of the assumed solution (the ansatz), where $\varphi$ and $\psi$ 
 are polynomials with undetermined coefficients. Equating the
 coefficients of the same powers of the exponentials or (and) equal powers
 of the self-similar variables, we obtain are overdetermined,
 in general, system of nonlinear algebraic equations for the unknown 
coefficients of the polynomials $\varphi$ and $\psi$ 
 and the phase constants. The solution of this system (if it exists) is 
necessary and sufficient for constructing an explicit expression for 
the solution of the original equation. We note that it was precisely 
an assumption about the form of the solutions of equations of the form (0.1)
 that made possible the successful use of constructive methods of solution 
finding, for example, modification of Hirota's method ([2], p. 184,[2'],p.68).

    In this paper, we describe a modification of the Painlev$\acute{e}$ 
 test for systems of semilinear parabolic equations. It turns on that
 the set of possible velocities of self-similar solutions of systems is,
 as in the case of individual equations, restricted and can be uniquely 
determined. The analogy with the scalar case made it possible to assume 
that the systems have solutions of  the definite form, for the finding of
 which the method of undetermined coefficients described above was used. 
  It is characteristic  that in the case of the systems considered in this 
paper, as in the one-dimensional case, it is, as a rule, possible to find
 all  similar solutions whose velocity is determined by the Painlev$\acute{e}$
 test by making a restriction to functions of the form (0.2). 
By means of the same method we have succeeded in constructing two-phase
 solutions of the form (0.3) of some of the consider systems.

We note however that the mechanism of interaction of waves that are 
solutions of systems of semilinear parabolic equation has been found 
to be very specific. Thus, in [4] (p. 44,[2']p.70) there is an asymptotic 
description of the process of wave creation from a finite perturbation 
for the Kolmogorov - Petrovskii - Piskunov (KPP) - Fisher equation,
 from which there follows the impossibility of constructing an expression 
that describes the interaction of the waves in the form dictated by 
Hirota's method. It turned out that the situation is quite different 
for systems of semilinear parabolic equations. 
In the paper, we consider example of a system for which explicit
 expressions for a solution describing the annihilation of kinks are 
constructed (In this text the figures are not reduced. Look figures 
in the original of a paper in TMF .
see Fig. 2 below).

The subject of the investigation in the present paper is systems of the 
form
\begin{eqnarray}
 U_t-U_{xx}=U^{l}(1-U^{m}-\theta^{n}) ,\nonumber\\
\theta_t-\theta_{xx}=-B U^{k}\theta^{q},
\label{5}
\end{eqnarray} 
where $U,\theta $  are the unknown functions,$ B $ is a nonvanishing 
constant, 
and $ m, n, k, q, l  $ are natural numbers. The paper consist of two 
sections. 
In Sec. 1, we give self-similar solutions of a system that is one of the
 forms of the Belousov - Zhabotinskii reaction model [8,9] and
has, in addition, 
 a biophysical application [10]. 
In Sec. 2, we give self-similar and two-phase solutions of various systems 
describing processes of the type of nonlinear kinetics.

\setcounter{equation}{0}

\section{SELF-SIMILAR SOLUTIONS\\ IN THE
BELOUSOV--ZHABOTINSKII\\ REACTION MODEL}
\label{s1}

Suppose that in (0.5)$ m = n=k-q = l=l, B\neq 0,B\neq 1/2 $, i.e., the system 
has the form
\begin{eqnarray}
 U_t-U_{xx}=U (1-U-\theta) ,\nonumber\\
\theta_t-\theta_{xx}=-B U \theta ,
\label{6}
\end{eqnarray} 
Then the following theorem holds.
\begin{them}
\label{th1}
All self-similar solutions of the system (1.1) that can be represented
 by a Laurent series in a punctured neighborhood of the moving 
pole $ x= -pt+c $, where $ p $ is the velocity and c is an arbitrary 
constant, have the form
\begin{eqnarray}
&& U=\frac{1}{(1+\exp{\tau})^{2}} ,~~~\theta=\frac{(1-B)(2\exp{\tau}+\exp{2\tau}}{(1+\exp{\tau})^{2}}, \nonumber\\
&&\tau=a x+b t ,~~~a^{2}=B/6,~~~b=-5B/6;
\label{7}
\end{eqnarray} 
\begin{eqnarray}
&& U=\frac{1-2\exp{\tau})}{(1+\exp{\tau})^{2}} ,~~~\theta=\frac{(1-B)\exp{2\tau}}{(1+\exp{\tau})^{2}}, \nonumber\\
&&\tau=a x+b t ,~~~a^{2}=-B/6,~~~b=-5B/6,~~~p=b/a;
\label{8}
\end{eqnarray} 
\end{them}

{\it Proof.}

 The fact that the given functions solve the system can be verified 
by substitution. We consider in more detal question of the uniqueness 
of the presented self-similar solutions associated with modification 
of the Painlev$\acute{e}$  test for system. We shall seek a solution 
in the form of Laurent series in powers of $\tau $.
\begin{eqnarray}
&& U=\sum_{k=-q_1}^{\infty} a_{k+q_1}(x+p t+c_1)^{k}, \nonumber\\
&& \theta=\sum_{k=-q_2}^{\infty} B_{k+q_2}(x+p t+c_1)^{k}, 
\label{9}
\end{eqnarray} 
where $ q_1 $ and $ q_2 $ are certain natural numbers that determine the
 order of the poles of the functions  $U $ and $\theta$, 
respectively, and in what follows we shall, without loss of generality,
 assume that the constants $c_1$ and $c_2$ are equal to zero. 
In the general case, the possible values of $ q_l$ and $ q_2$ are 
determined by means of Newton's polygon method; however, in the present
 case it can be done by using elementary arguments and equating 
the exponents of the definitely highest negative powers of the second 
derivatives and the quadratic terms of the polynomials.
 It is easy to see that in the given case $ q_1 =q_2 = 2$, i.e.,
 both functions $ U $ and $ \theta $ have a pole of second order. 
Substituting the obtained series in the system and equating the 
coefficients of equal powers of $ \tau $, we obtain a system of nonlinear 
algebraic equations for the determination of the coefficients of the series:
\begin{eqnarray*}
&&a_0(a_0 + b_0 - 6) = 0,~~~ b_0(a_0 B - 6) = 0,\nonumber\\
&&2a_1 a_0 +a_1 b_0 -2a_1 + a_0 b_1 - 2a_0 p = 0,\nonumber\\
&&a_1 b_0 B+a_0 b_1B - 2b_1 - 2b_0 p = 0,\nonumber\\
&& 2a_2 a_0 + a_2 b_0 + {a_1}^{2} + a_1 b_1 - a_1 p + a_0 b_2 - a_0 = 0,
\nonumber\\
&&a_2 b_0 B+ a_1 b_1 B + a_0 b_2B - b_1 p - 0,\nonumber\\ 
&&2a_3 a_0 + a_3 b_0 + 2a_2 a_1 + a_2 b_1 +a_1b_2 - a_1 + a_0 b_3 = 0,\nonumber\\
&&B(a_3b_0 + a_2b_1 + a_1b_2 + a_0b_3) = 0,\nonumber\\
\end{eqnarray*}
\begin{eqnarray}
&&2a_4 a_0 + a_4 b_0 - 2a_4 + 2a_3 a_1 +a_3 b_1 + a_3 p+a_2^{2}\nonumber\\  
&&\qquad + a_2 b_2 - a2 + a_1 b_3 + a_0 b_4 - 0,\nonumber\\
&&a_4 b_0B + a_3 b_1 B + a_2 b_2B + a_1 b_3B + a_0 b_4B - 2 b_4 + b_3 p = 0,\nonumber\\
&&2 a_2 a_0+ a_5 b_0 - 6 a_5 + 2a_4 a_1 + a_4 b_1 + 2a_4 p + 2a_3 b_2\nonumber\\  
&&\qquad 
+  a_3 b_2- a_3 + a_2 b_3 + a_1 b_4 + a_0 b_5 = 0,\nonumber\\
&&a_5 b_0 B+ a_4 b_1 B + a_3 b_2 B + a_2 b_3 B + a_1 b_4 B + a_0 b_5 B - 
 6 b_5 + 2 b_4 p = 0, \nonumber\\
&&2 a_6 a_0+ a_6 b_0 - 12a_6 + 2 a_5 a_1 + a_5 b_1 + 3 a_5 p + 2a_4 a_2\nonumber\\
&&\qquad  
+ a_4 b_2- a4 + {a_3}^2 + a_3 b_3 + a_2 b_4 + a_1 b_5 + a_0 b_6 = 0,\nonumber\\
&&a_6 b_0 B+ a_5 b_1 B + a_4 b_2 B + a_3 b_3 B + a_2 b_4 B + a_1 b_5 B\nonumber\\
&&\qquad + a_0 b_6 B -12 b_6  + 3 b_5 p = 0. 
\label{10}
\end{eqnarray} 

Solving successively this system for the unknown coefficients $ a_{k} $
 and  $ b_{k}$, we obtain
\begin{eqnarray}
&&a_0=\frac{6}{B},~~~b_0=\frac{6(B-1)}{B},~~~ a_1= \frac{6p}{5 B},~~~b_1=\frac{6(B-1)p}{5 B},\nonumber\\
&&a_2=\frac{25B-p^2}{50 B},~~~b_2 = -\frac{(1-B)(p^2+25 B)}{50 B},\nonumber\\
&&a_3=\frac{p^3}{250B},~~~b_3=\frac{(B-1)p^3}{250 B},\nonumber\\
&&a_4 =\frac{125 B^{2}-7 p^{4}}{5000 B},   ~~~    b_4 =\frac{(B-1)(125 B^{2}-7 p^{4})}{5000 B},   \nonumber\\
 &&a_5 =-\frac{1375 B^{2}-79 p^{4}}{75000 B},   ~~~    b_5 =\frac{p(1-B)(1375 B^{2}-79 p^{4})}{75000 B},   
\nonumber\\
 && a_6 =\frac{37500 b_6-625 B^{2}p^{2}+36 p^{4}}{37500(B-1)}.
\label{11}
\end{eqnarray} 

It is readily seen that in the final equation of this system the 
coefficient of the unknown $b_6$ is equal to zero and the obtained equation 
is an expression from which the possible values of the velocity can be
 determined. This expression is usually called the dispersion relation. 
It has the form
\begin{eqnarray}
(2B-l)(36p^{4} -625B^{2}) = 0.
\label{12}
\end{eqnarray} 	
By virtue of the conditions of the theorem, the solution of $Eq. (1.6!!)$
 for p gives two (up to the sign) possible values of the
\begin{eqnarray}
p_{1,2}=\frac{\pm 5 \sqrt{B}}{\sqrt{6}},~~~ p_{3,4}=\frac{\pm 5 i \sqrt{B}}{\sqrt{6}} .
\label{13}
\end{eqnarray} 	
For each particular value of the velocity $ p $ , the coefficients $ a_k  $ 
and $ b_k $ of the series for $U $ and  $\theta$ are uniquely determined;
 therefore, there exist only two different self-similar solutions of 
the system that can be represented by a Laurent series with second-order
 pole. It is obvious that to each of these solutions there corresponds 
one of the two possible values of the velocity and that for a definite 
value of the parameter $ B $ with fixed sign these values are different.
 By virtue of the invariance of th substitution $ x->- x $ in the system 
(1.5), reversal of the sign of the phase constant $ a $ corresponds to 
reflection of the traveling wave with respect to the ordinate with
 change of the direction of its motion, i.e., reversal of the velocity. 
Thus, the above solutions completely exhaust the store of velocities 
obtained by means of the Painlev$\acute{e}$   test.
$\square$
\bigskip

\begin{no}
\label{no1}
\rm
 It is clear that for any $ B\neq0 $ only one of the two solutions (1.2) 
and (1.3) is real. The boundary condition that real solutions of the 
system (1.1) satisfy have the form
\begin{eqnarray}
U_{x\rightarrow\infty}\rightarrow0,~~~\theta_{x\rightarrow\infty}\rightarrow1-B,
~~~U_{x\rightarrow-\infty}\rightarrow1,~~~\theta_{x\rightarrow-\infty}\rightarrow0.
\label{14}
\end{eqnarray} 	
\end{no}
\begin{no}
\label{no2}
\rm
  It follows from $ Eq. (1 .6)$ that for $ B= l/2 $ there may also exist
 other solutions in addition to (1 .2) and (1 .3) that are given 
by explicit
 expressions.
\end{no}

Note that as $ B\rightarrow1 $ in each of the solutions (1.2) and (1.3)
$\theta\rightarrow0$, and at the same time the solution of the system it
 continuously transformed into the solution of the scalar equation
\begin{eqnarray}
U_t-U_{xx}-U(1-U) = 0.
\label{15}
\end{eqnarray} 
which is known as the KPP - Fisher equation.  The Painlev$\acute{e}$  
test for Eq. (1.8) gives, up to the sign, two possible values of that
 velocity for solutions that can be represented by a Laurent series,
 namely
\begin{eqnarray}
p_{1,2}=\frac{\pm 5 }{\sqrt{6}},~~~ p_{3,4}=\frac{\pm 5 i }{\sqrt{6}} .
\label{16}
\end{eqnarray} 
The solutions with such values of speed were earlier calculated in [2] p.184.

  It is therefore obvious that there do not exist any other solutions of
 Eq. (1.8) that can be expanded in a Laurent series apart from those 
that are obtained from the expressions (1.2) and (1.3) for B= 1 .
 In addition, it is clear that in the given case the velocity (1.7)
 are continuous functions of the parameter $ B$ of the system on 
the complete domain of definition, i.e., for any $ B\neq0 $ will be 
shown below that this is not always the case.

\setcounter{equation}{0}

\section{SELF-SIMILAR AND TWO-PHASE SOLUTIONS IN MODELS OF
NONLINEAR KINETICS}
\label{s2}

Suppose that in (0.5) $ m=n=k=2, q=l= 1 $.  We obtain the system
\begin{eqnarray}
 U_t-U_{xx}=U (1-U^{2}-\theta^{2}) ,\nonumber\\
\theta_t-\theta_{xx}=-B U^{2}\theta,
\label{17}
\end{eqnarray} 
As in the case of the system (1.1), the order of the poles of the 
functions  $ U $ and $\theta$ in the system (2.1) is determined.   In the
given case, $ ql = q2 = 1 $ in (1.4) and (1.5). 
The unique (up to sign) possible expression for the velocity that is
 obtained by me;
of the Painlev$\acute{e}$  test has the form
\begin{eqnarray}
p_{1,2}=\frac{\pm 3 \sqrt{B}}{\sqrt{2(2 B-1)}} .
\label{18}
\end{eqnarray} 	
It proved to be possible to construct solutions of the system (2. 1) for 
$B=2 $.
\begin{them}
\label{th2}
All self-similar solutions of the system (2.1), $B=2$ , represented by 
a Laurent series in the punctured neighborhood of the moving pole $ x= -pt+c $ ,
 where $ p $ is the velocity, and c is an arbitrary constant, 
have the form
\begin{eqnarray}
&& U=\frac{\pm1}{1+\exp{\tau}} ,~~~\theta=\frac{\pm\exp{\tau}}{1+\exp{\tau}}, \nonumber\\
&&\tau=a x+b t ,~~~a=\pm1,~~~b=-1,~~~p=b/a=\pm1;
\label{19}
\end{eqnarray} 
\begin{eqnarray}
&& U=\frac{-a}{a-\tau} ,~~~\theta=\frac{\pm\exp{\tau}}{1+\exp{\tau}}, \nonumber\\
&&\tau=a x+b t ,~~~b=2 a,~~~p=b/a=2;
\label{20}
\end{eqnarray} 
\end{them}
where $c$ is an arbitrary constant.
{\it Proof.}
 The fact that functions (2.3) and (2.4) satisfy the system can be 
verified by substitution. The Painlev$\acute{e}$ test for the considered
 system gives two possible values of the velocity for solutions that
 can be represented by a Laurent series:  $ p_1=2, p_2=\pm1 $. 
It is with such velocities that the above solutions move.
 From this the statement of the theorem follows, and the theorem is proved.
$\square$
\bigskip

\begin{no}
\label{no3}
\rm
 The value of the expression (2.2) for $ B=2 (p=\sqrt{3})$ is different
 from the values of the velocity given in Theorem 2 for the solutions of 
the system (2.1) for the same value of the parameter $B$. 
Thus, the expression (2.2) is not a continuous function of the
parameter~$B$ of the system (2.1).  
\end{no}

{\bf Proposition 1. }  The system (2.1), $B = 2$, has a two-phase 
(nonself-similar) solution of the form
\begin{eqnarray}
&& U=\frac{a_1+\exp{\tau_2}}{a_1+\tau_1+\exp{\tau_2}} ,
~~~\theta=\frac{\pm\tau_1}{a_1+\tau_1+\exp{\tau_2}} , \nonumber\\
&&\tau_1=a_1 x+b_1 t ,~~~\tau_2=a_2 x+b_2 t ,\nonumber\\
&&b_1=-2 a_1,~~~b_2=1,~~~a_2=1;
\label{21}
\end{eqnarray}
where $ a_1 $ is an arbitrary constant.
The proposition is verified by substitution. The fnteraction of the
 self-similar waves (2.3) and (2.4) described by (2.5) is analogous 
to the interaction of a kink with a discontinuous solution, 
which can be described by explicit expressions for individual equations
 with cubic nonlinearity [4] (p. 39-42),[2']p.51. Figure 1 gives the graph 
of the function $ U $ for $a_1 = 1$. It can be seen that on the collision
 of the singularity with the front of the kink the order of the latter 
increases. For $a_l = - 1 $ , the interaction of the discontinuous 
solution with the kink described by the function $ U $ leads to 
annihilation of the singularity and the formation of one kink.
 A situation analogous to the considered cases also holds for the
 function $ \theta $.
We turn to the next example.   Suppose that in (5) $ m=n = q=l, l=k=2 $. 
  The resulting system has the form
\begin{eqnarray}
 U_t-U_{xx}=U^{2}(1-U-\theta) ,\nonumber\\
\theta_t-\theta_{xx}=-B U^{2}\theta,
\label{22}
\end{eqnarray} 
In this case, it proved possible to construct both self-similar and 
two-phase solutions containing the parameter $ B $ of the original
 system (2.6) as a free parameter.
The Painlev$\acute{e}$  test for the system (2.6) gives two possible 
values of the velocity for self-similar solutions that can be represented 
by a Laurent series, namely
\begin{eqnarray}
p_{1}=\frac{\pm \sqrt{B}}{\sqrt{2}} ~~~p_{2}=\pm \sqrt{2B} .
\label{23}
\end{eqnarray} 	
  Note that this is true for any $ B\neq1/3 $, since for $ B= 1/3 $ 
the dispersion relation vanishes identically.   A consequence of this 
result is the following theorem, the proof of which is completely analogous
 to that of Theorems 1 and 2.
\begin{them}
\label{th3}
All self-similar solutions of the system (2.6), $ B\neq1/3 $, that can 
be represented by a Laurent series in the punctured neighborhood of 
the moving pole$ x= -pt+c$, where p is the velocity and c is an arbitrary 
constant, have the form
\begin{eqnarray}
&& U=\frac{c_1}{a_1+a_2 \exp{\tau}} ,\theta=\frac{(1-B) a_2 \exp{\tau}}{c_1+a_2\exp{\tau}}, \nonumber\\
&&\tau=a x+b t ,~~~a^{2}=B/2,~~~b=-B/2,~~~p=b/a=\pm\frac{\sqrt{B}}{\sqrt{2}},
\label{24}
\end{eqnarray} 
$c_1 , c_2 $ are arbitrary constants;
\begin{eqnarray}
&& U=\frac{a\sqrt{2}}{\tau\sqrt{B}+a\sqrt{2}} ,\theta=\frac{\sqrt{B}(1-B)\tau}{\tau\sqrt{B}+a\sqrt{2}} , \nonumber\\
&&\tau=a x+b t ,~~~b=- a\sqrt{2B},~~~p=-\sqrt{2B},
\label{25}
\end{eqnarray} 
$ a $ is an arbitrary constant.
\end{them}
\begin{no}
\label{no4}
\rm
 As in the case (1.1), for $ B=1/3 $ there can be other solutions
 in addition to (2.8) and (2.9) that can be represented by explicit 
expressions.

\end{no}

{\bf Proposition 2}.  The system (2.6) has a two-phase solution of the 
form
\begin{eqnarray}
&& U=\frac{a_1+a_2\exp{\tau_2}}{a_1+a_2(\tau_1+\exp{\tau_2}} ,
~~\theta=\frac{a_2(1-B)\tau_1}{a_1+a_2(\tau_1+\exp{\tau_2}} , \nonumber\\
&&\tau_1=a_1 x+b_1 t ,~~~\tau_2=a_2 x+b_2 t ,\nonumber\\
&&b_1=- 2a_1 a_2,~~~a_2^{2}=B/2,~~~b_2=B/2,
\label{26}
\end{eqnarray} 
 is $a_1$ arbitrary constant.	
The validity of the proposition is proved by substituting the 
expressions (2.10) in the system (2.6).
As in the case of (1.1), the self-similar solutions of the system (2.6), 
and also the two-phase solution (2.10) describing their interaction
 are reduced continuously in the limit $B\rightarrow1$, respectively, 
to the self-similar and two-phase solutions of the one-dimensional 
problem known as one of the forms of the Zel'dovich equation 
([2], p. 73, p. 198): $ U_t-U_{xx}-U^{2}(1 - U)=0$. Here, as in the
 system (1.1), the velocities (2.7) are continuous functions of the 
parameter B on the complete domain of definition. Note that all the 
given solutions of the system (2.6) are real for $ B>0 $.
The functions (2.10) describe the interaction of self-similar solutions
 of the system (2.6). The nature of this interaction is similar to the 
case considered above (see Fig. 1 in the original of a paper in TMF ).
We now give an example of the system (5) in which it proved possible 
to find not only self-similar solutions but also a bounded two-phase
 solution that is the result of interaction of waves propagating 
in strips.

 Suppose that in (5) $ m=k=2, q = n = l=l $. Then we obtain the system 
\begin{eqnarray}
 U_t-U_{xx}=U(1-U^{2}-\theta) ,\nonumber\\
\theta_t-\theta_{xx}=-B U^{2}\theta,
\label{27}
\end{eqnarray} 
\begin{them}
\label{th4}
The system (2.11) has nontrivial self-similar solutions that can be 
represented by a Laurent series for a unique value of the parameter $B$,
 namely,$ B=1$.
\end{them}

{\it  Proof.} Functions $U$ and $\theta$ that can be represented 
by a Laurent  
series in powers of t and satisfy the system (2.11) have a first-order
 pole, i.e., the coefficients of $a$ and $b_0$ must be nonzero. It follows
 from the system of equations for these coefficients that this is possible
 only if $B = 1$:
\begin{equation}
a_0(a_0^{2}-2)=0 ,~~b_0(a_0^{2}B -2 )=0. 
\label{28}
\end{equation} 

The theorem is proved.
$\square$
\bigskip

{\bf Proposition 3}.  There exist self-similar solutions of the system
 (2.11), $ B=1 $, having the form
\begin{eqnarray}
&& U=\frac{\pm1}{1+\exp{\tau}} ,
~~\theta=\frac{\exp{\tau_1}}{1+\exp{\tau_2}} , \nonumber\\
&&\tau=a x+b t ,~~ ~b=-1/ 2,~~~a^{2}=1/2,
\label{29}
\end{eqnarray} 

{\bf  Proposition 4. } There exists a two-phase solution of the system 
(2.11), $ B=1 $, having the form
\begin{eqnarray}
&& U=\frac{1-\exp{\tau_1}}{1+\exp{\tau_1}+\exp{\tau_2}} ,
~~\theta=\frac{\exp{\tau_2}}{1+\exp{\tau_1}+\exp{\tau_2}} , \nonumber\\
&&\tau_1=a_1 x+b_1 t ,~~~\tau_2=a_2 x+b_2 t ,\nonumber\\
&&b_1=0,~~~a_1=\sqrt{2}, ~~~a_2=1/\sqrt{2},~~~b_2=-1/2,
\label{30}
\end{eqnarray} 

$\square$
\bigskip

The validity of both propositions can be proved by substituting the 
given solutions in the system (2.11).
The two-phase functions $U $ and $ B $ are the result of interaction 
of the self-similar waves described in Proposition 3. Graphs of the 
two-phase solution are shown in Fig. 2. It can be clearly seen that
 the interaction described by the function 6 (see Fig. 2a) represents 
the annihilation of kinks propagating in the strip between $ 0 $ and $1 $
(for $t>t_0 $, the solution is close to zero). In the case of the function 
$U $(see Fig. 2b), there is an interaction of the type of merging of
waves propagating in different strips [for $ t>t_0 $, a single wave
 close to the stationary solution-of Semenov's equation ([2], p. 193)
 is formed].

\end{sloppypar}

\end{document}